\documentclass[journal,10pt]{IEEEtran}
\makeatletter
\def\endthebibliography{%
	\def\@noitemerr{\@latex@warning{Empty `thebibliography' environment}}%
	\endlist
}
\makeatother

\newcommand{\RomanNumeralCaps}[1]
{\MakeUppercase{\romannumeral #1}}

\usepackage{cite}

\usepackage[pdftex]{graphicx}

\usepackage{filecontents}
\usepackage{amssymb}
\usepackage{color}

\usepackage{epsfig}

\usepackage[cmex10]{amsmath}

\usepackage{url}

\usepackage{algpseudocode}
\usepackage{algorithm}

\usepackage{subcaption}

\usepackage{lipsum}

\usepackage{siunitx}

\usepackage{soul,color}

\usepackage{mathrsfs}

\usepackage{array}

\usepackage[inline]{enumitem}

\usepackage{breqn}

\usepackage{multirow}
\usepackage{array}
\newcolumntype{L}[1]{>{\raggedright\let\newline\\\arraybackslash\hspace{0pt}}m{#1}}
\newcolumntype{C}[1]{>{\centering\let\newline\\\arraybackslash\hspace{0pt}}m{#1}}
\newcolumntype{R}[1]{>{\raggedleft\let\newline\\\arraybackslash\hspace{0pt}}m{#1}}

\begin{document}
	
	\title{\LARGE Energy-efficient Deployment of Multiple UAVs\\Using Ellipse Clustering to Establish Base Stations}

	\author{Si-Chan Noh,~\IEEEmembership{Member,~IEEE}, Hong-Bae Jeon,~\IEEEmembership{Student Member,~IEEE}, and \\Chan-Byoung Chae,~\IEEEmembership{Senior Member,~IEEE}
		\thanks{This work was supported by the Future Combat System Network Technology Research Center Program of Defense Acquisition Program Administration and Agency for Defense Development (UD160070BD). (\textit{Si-Chan Noh and Hong-Bae Jeon are co-first authors.}) (\textit{Corresponding author: Chan-Byoung Chae.})}
		\thanks{S.-C Noh was with the School of Integrated Technology, Yonsei University, Seoul 120-749, Korea. He is now with the Newratek, Inc., Seoul 06175, Korea (e-mail: sc.noh@newratek.com).}
		\thanks{H.-B. Jeon and C.-B. Chae are with the School of Integrated Technology, Yonsei University, Seoul 120-749, Korea (e-mail: \{hongbae08, cbchae\}@yonsei.ac.kr).}
		\thanks{Manuscript received XXX, XX, 2019; revised XXX, XX, 2020.}}
	
	\markboth{}
	{}

	\maketitle
	
	\begin{abstract}
		
		The demand for future wireless communication systems is being satisfied for various circumstances through unmanned aerial vehicles (UAVs), which act as flying base stations (BSs). In this letter, we propose an ellipse clustering algorithm that maximizes the user coverage probability of UAV-BSs and avoids inter-cell interference with minimal transmit power. We obtain the coverage of each UAV by adjusting its antenna half-power beamwidth,  orientation, and 3D location by minimizing the path loss of the cell-edge user. Simulation results confirm that the proposed algorithm achieves high system throughput and coverage probability with lower transmit power compared to conventional algorithms.

	\end{abstract}


	\IEEEpeerreviewmaketitle

	\section{Introduction}
    In wireless communication systems, several applications are increasingly using unmanned aerial vehicles (UAVs). Especially, the use of UAVs in cellular networks could be a solution in emergencies when base stations (BSs) are malfunctioning. For instance, during the Olympics, there is a considerable increase in the number of active users. Here, UAVs can act as a flying BSs to support existing terrestrial BSs by providing adequate user coverage~\cite{d2dmagazine}. Even during disasters, when existing BSs cannot provide communication links, UAV-BSs could serve as a public access point to enable wireless communications~\cite{RIICC,mozaeff}. 

Unlike terrestrial BSs, UAV-BSs can provide fast deployment to service-specific regions where communication links are not ensured~\cite{LTEUAV}. Furthermore, the high altitude of UAV-BSs can offer superior line-of-sight (LoS) links between the UAV and the users~\cite{a2gglobecom}. However, to fully exploit UAV-BSs, several technical limitations should be addressed. Neighboring UAV-BSs that interrupt user communication, for example, should be prevented with interference management. In addition, UAV-BSs possess limited onboard energy; this should be carefully managed so that users can be offered long-term services~\cite{arxv}. These issues can be solved by the 3D deployment and considering a practical and realistic radiation pattern of UAV-BSs. It is because the path loss between the UAV-BS and the user, which determines the received and interference power of the user, is determined by their 3D locations~\cite{lapotp, a2gglobecom}. Moreover, considering a practical and realistic radiation pattern prevents the undesired interference from other UAV-BSs and wasting unnecessary transmit power, which leads to an efficient usage of the UAV-BSs' onboard energy~\cite{battery}.

Several studies have addressed the 3D deployment and coverage problems associated with UAV-BSs~\cite{arxv}. 
The authors in~\cite{lapotp} analyzed the optimal altitude for a single UAV, determining that the UAV altitude is related to the maximum allowed path loss and parameters of the given environment, as defined by the International Telecommunication Union (ITU). The authors in~\cite{eemrtp} derived the optimal UAV deployment that minimizes energy consumption and guarantees quality-of-service (QoS) to users. Although the methods are limited to a single UAV downlink scenario.
In~\cite{mozaeff}, the authors proposed a coverage function that reflects the antenna gain and maximized the total coverage probability with multiple UAVs according to the circle packing strategy. The authors in~\cite{ruioptbeam} studied throughput optimization by jointly determining the UAV flying altitude and antenna half-power beamwidth (HPBW). However, these works assumed equal HPBWs for azimuth and elevation sides; this assumption ignores the practical antenna radiation beam pattern.

In this letter, we consider a multiple-UAV scenario, where each UAV is equipped with a directional antenna that provides elliptic coverage to ground users. The elliptic coverage reflects the practical radiation beam pattern. We group the users into non-overlapping elliptical regions to avoid inter-cell interference; this grouping is done according to our proposed ellipse clustering algorithm. We appropriately adjust the orientation of UAV and the HPBWs of the antenna to cover the users by elliptical coverage. We also propose the energy-efficient 3D deployment of UAVs that minimizes the total transmit power of UAVs while guaranteeing ensuring QoS for every user. 
	
	\section{System Model and Problem Formulation}
	
\begin{figure}[t]
		\begin{center}
			{\includegraphics[width=0.8\columnwidth,keepaspectratio]
				{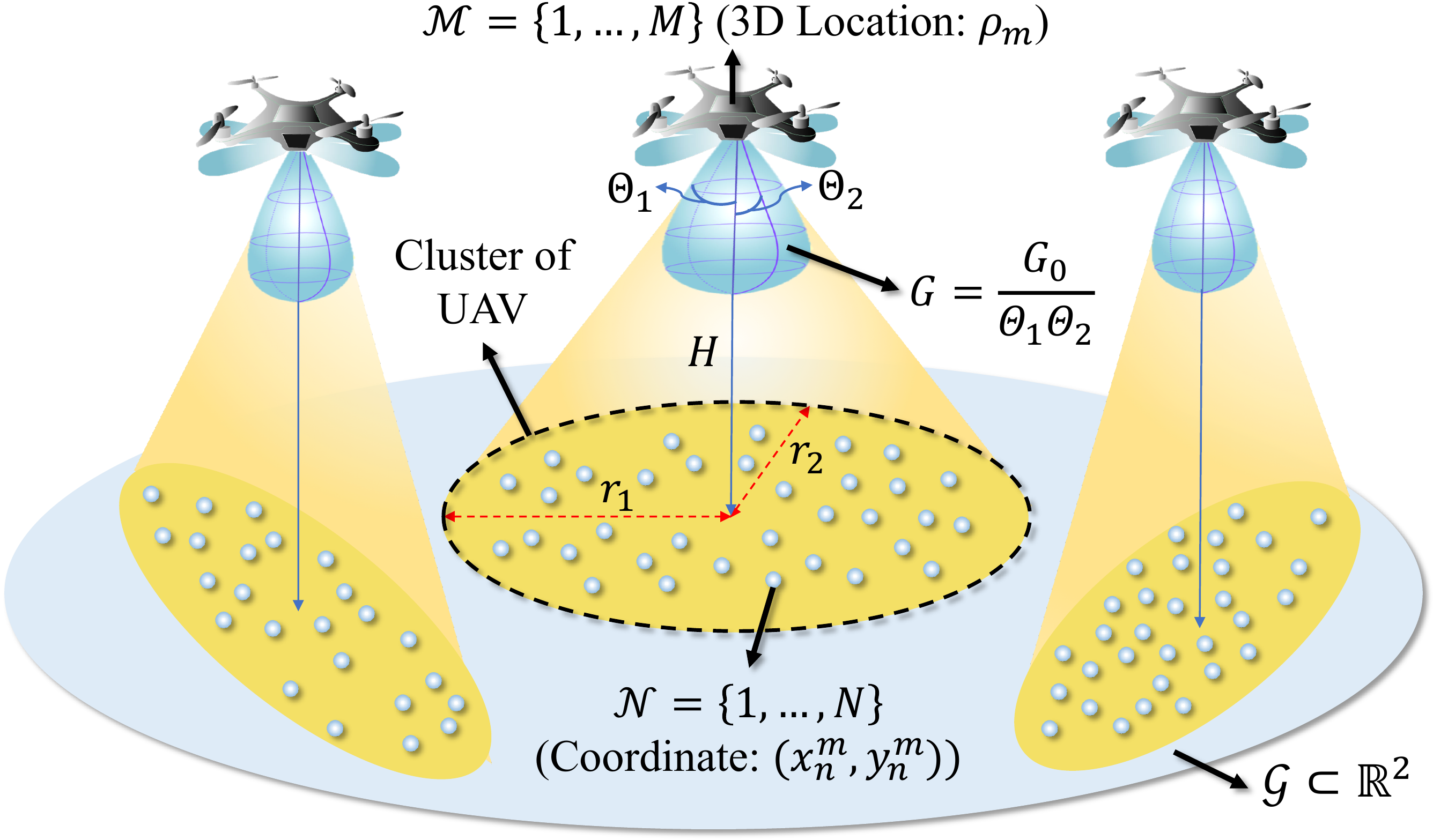}%
				\label{fig_polar_plot_25}}
			\caption{Downlink between multiple UAVs and ground users.}
			\label{fig_system_model}
		\end{center}
	\end{figure}

\begin{figure}[t]
		\begin{center}
			{\includegraphics[width=0.99\columnwidth,keepaspectratio]
				{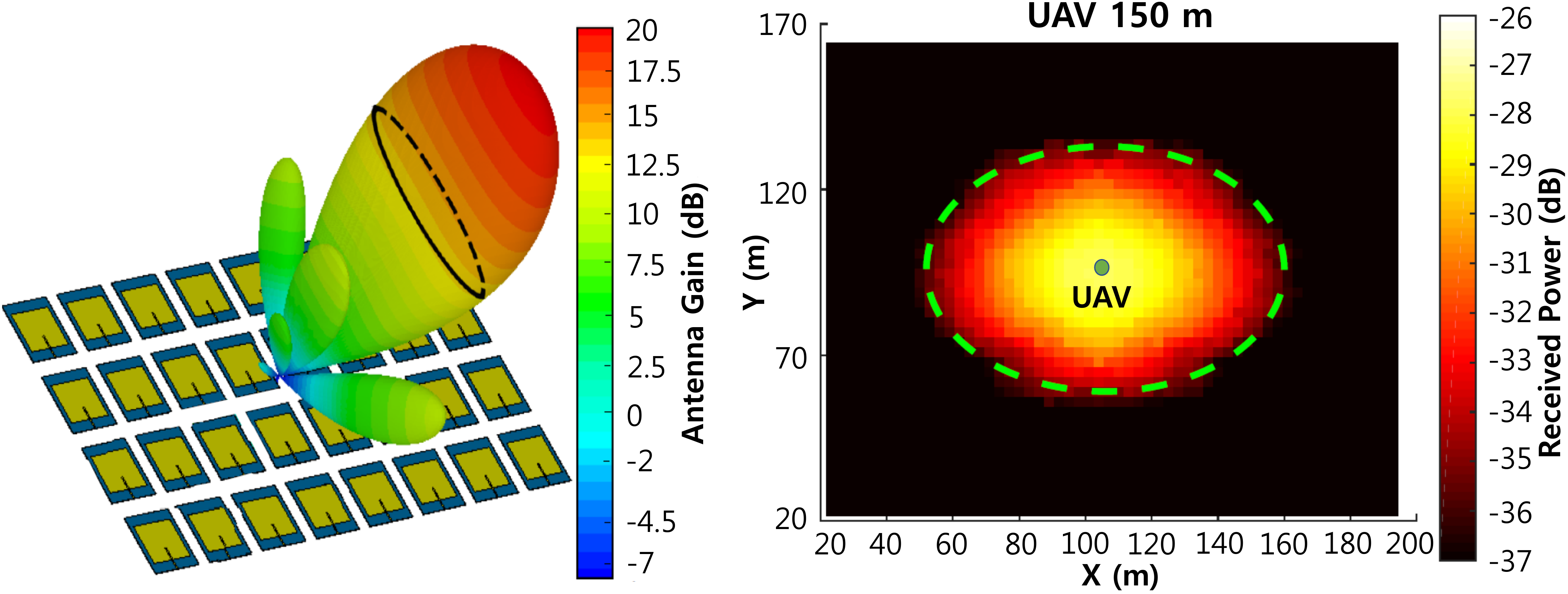}%
				\label{beam}}
			\caption{Directional antenna beam pattern.}
			\label{fig_beam}
		\end{center}
	\end{figure}

	As illustrated in Fig.~\ref{fig_system_model}, we consider the geographical area $\mathcal{G}$ $\subset$ $\mathbb{R}^2$ containing $N$ users arbitrarily distributed according to the distribution $f(x,y)$. We deploy UAV-BSs to provide wireless service for ground users in the downlink. Let the index sets of users and UAVs be $\mathcal{N}$ = $\left\{ {1,~...,~N }\right\}$ and $\mathcal{M}$ = $\left\{ {1,~...,~M }\right\}$, respectively, where $M$ is the number of UAVs and it is determined by $N$ and the distribution of the users. We assume that each UAV covers a different user cluster through stationary hovering above the center of the cluster without rolling or pitching and that the cluster area is not affected by the users in adjacent clusters. Furthermore, we denote the coordinates of user $n$ $\in$ $\mathcal{N}$ covered by UAV $m$ by  $(x_n^m,y_n^m)$, and the 3D location of UAV $m$ by $\rho_m\in \mathbb{R}^3$. Each UAV-BS is equipped with a directional antenna with adjustable beamwidth, and each user is assumed to be equipped with a unit-gain omnidirectional antenna. In addition, we assume that the azimuth and elevation HPBWs of the UAV directional antenna are not equal, and they are denoted as $2\Theta_1$, $2\Theta_2$ $\in$ $(0,{\pi})$, respectively. Thus, the antenna gain $G$ along the azimuth $\theta$ and elevation $\phi$ can be approximated as~\cite{beamwidth} 
	\begin{align}
	\label{antennagain}
	\ {G} = \left\{\begin{array}{cl}
	\frac{G_0}{\Theta_1 \Theta_2} & (-\Theta_1 \le \theta \le \Theta_1 ,-\Theta_2 \le \phi \le \Theta_2) \\
	0 & \ $(otherwise)$, \end{array} \right. \
	\end{align} 
	where $G_0$ $\approx$ 30,000, $\theta$ and $\phi$ are given in degrees, and zero sidelobe gain is assumed. An example of the 3D radiation pattern of an $8 \times 4$ patch antenna and the measured power received by the users are illustrated in Fig.~\ref{fig_beam}. We set the transmit power to 40~dBm using a directional antenna at an altitude of 150~m. The received power was also measured and computed by using Wireless System Engineering (WiSE), a 3D ray-tracing tool developed by the Bell Labs. In Fig.~\ref{fig_beam}, we can see that the feature of the main lobe is elliptical, which implies different azimuth and elevation HPBWs. By assuming that the UAV is at the center of the beam pattern, the coverage by the antenna main lobe is determined by the maximum distance $r_1$ $=$ $H$ $\tan(\Theta_1)$, where $H$ is the height of UAV, and the minimum distance  $r_2$ $=$ $H$ $\tan(\Theta_2)$, provided that ${\Theta_1}$ $\geq$ ${\Theta_2}$.

	Next, to model the air-to-ground channel, we consider the $\text{LoS}$ and non-LoS (NLoS) components between the UAV and the ground users. The probability of $\text{LoS}$ components depends on the elevation and relative position between the UAV and the users. For example, the characteristics of the environment affected by the surrounding buildings also determine the probability. From~\cite{a2gglobecom}, the LoS probability $P_{n,\text{LoS}}^{(m)}$ of user $n$ covered by UAV $m$ is modeled by the function of the altitude $h_m$ of UAV $m$ and the horizontal distance $r_n^{(m)}$ between the UAV $m$ and the user $n$. Further, the path loss $PL^{(m)}_{n}$ of the \text{LoS} and \text{NLoS} links from UAV $m$ for user $n$ in dB can be expressed as
	\begin{align}
	\label{Pathloss}
	\text{\it PL}^{(m)}_{n} = \left\{ \begin{array}{cl}
	\text{FSPL}_{n}^{(m)} + \mathcal{E}_{\text{\scriptsize LoS}} - G_{m} & $(LoS link)$ \\
	{\:}{\:}{\:} \text{FSPL}_{n}^{(m)} + \mathcal{E}_{\text{\scriptsize NLoS}}  - G_{m} &  $(NLos link)$, \end{array} \right. 
	\end{align} 
	where $\text{FSPL}_{n}^{(m)}$ is the free-space path loss of the UAV $m$ covering user $n$ in dB. It is a function of the distance $d_{n}^{(m)}$ $=$ $\sqrt{(r_{n}^{(m)})^2 +(h_m)^2}$ between the UAV $m$ and covered user $n$. Variables $\mathcal{E}_{\text{\scriptsize LoS}}$ and $\mathcal{E}_{\text{\scriptsize NLoS}}$ are the excessive path losses in dB that depend on the environment type, such as suburban, urban, dense urban, and high-rise urban, as detailed in~\cite{a2gglobecom}. Variable $G_{m}$ is the antenna gain of UAV $m$ in dB. Finally, the average path loss $	\overline{{ PL}}^{(m)}_{n} $ between UAV $m$ and user $n$ is derived as
	\begin{align}
	\label{avgpathloss}
	\overline{{ PL}}^{(m)}_{n}  =  10^{\frac{\text{\scriptsize{FSPL}}_{n}^{(m)}-{\text{\it \scriptsize G}_{m}}}{10}} \left(P^{(m)}_{n,\text{LoS}}10^{
		\frac{\mathcal{E}_{\text{\tiny LoS}}}{10}}+  {P^{(m)}_{n,\text{NLoS}}}10^{\frac{\mathcal{E}_{\text{\tiny NLoS}}}{10}}  \right).
	\end{align}
	Note that $	\overline{{ PL}}^{(m)}_{n} $ is a function of $h_m$ and $r_{n}^{(m)}$.
	
	To avoid inter-cell interference considering the practical 3D antenna beam pattern, we aim to optimize groups such that no user belongs to multiple clusters. In this situation, to guarantee the QoS, the signal-to-noise ratio (SNR) $\zeta^{(m)}_{n} = \frac{P^{(m)}_{r,n}}{\sigma^2}$ of user $n$ covered by UAV $m$ must be greater than the minimum SNR threshold $\zeta_{\text{th}}$. Here, $\sigma^2$ is the noise power, and $P^{(m)}_{r,n} = \frac{P_{t,m} }{\overline{PL}^{(m)}_{n}}$ is the power received at user $n$ from UAV $m$, where $P_{t,m}$ is the transmit power of UAV $m$.
	
Given the non-overlapping user-UAV association condition and the user-specific SNR constraints, our aim is to minimize the total transmit power of the UAVs to achieve energy efficiency, which is achieved by adjusting the number and 3D location of the UAVs and the antenna HPBW. This optimization problem can be expressed as
	\begin{equation}
	\label{object}
	\begin{aligned}
	& \underset{\ \{C_m\},\{\rho_m\},M}{\text{min}}
	& & \sum\limits_{m=1}^M P_{t,m}{\:}{\:} \\
	& \text{~~~~~~~s.t.}
	& & C_m  \cap  C_{m'} = \varnothing {\:}{\:} (\forall m \neq m', {\:}{\:} m,~m' \in \mathcal{M}), \\
	&&& \sum\limits_{m=1}^M |C_m| = N,  ~\frac{P^{(m)}_{r,n}}{\sigma^2} \geqslant \zeta_{\text{th}},\\
	&&& h_{\textrm{min}}\le h_m \le h_{\textrm{max}}~(\forall m \in  \mathcal{M},\forall n \in  C_m),
	\end{aligned} 
	\end{equation}
	where $C_m\subset{\mathcal{N}}$ is the set of users assigned to UAV $m$. The parameters $h_{\textrm{min}}$ and $h_{\textrm{max}}$ are the lower and upper bounds of the UAV's altitude, respectively. In this letter, we set $h_{\textrm{min}}=r_1\tan (\frac{\pi}{12})$, where $\frac{\pi}{12}$ is the lowest elevation angle $\theta=\tan^{-1} (\frac{h_m}{r_n^{(m)}})$ of UAV that leads to valid $P_{n,\text{LoS}}^{(m)}$ we use~\cite{a2gglobecom}. Therefore, to solve~(\ref{object}), we should find $M$ and $\{C_m\}_{m=1}^M$ that prevents inter-cell interference based on the QoS condition. However,~(\ref{object}) is generally difficult to solve given the high number of unknowns and nonlinear constraints, and it is regarded as an NP-hard problem. To reduce the complexity, we perform an ellipse clustering for the users, as detailed in Section~\RomanNumeralCaps{3}.

	
	\section{Energy-Efficient Ellipse Clustering}
	
	We propose an ellipse-clustering algorithm based on different azimuth and elevation HPBWs for UAV deployment to conform UAV-BSs. We assume that the antenna of the UAV is vertically tilted toward the ground. Therefore, adjusting the antenna’s HPBWs and 3D location as well as the orientation of the UAV allows us to cover the desired area without interference from the users.
	
	
	\subsection{Hierarchical Clustering Based on Silhouette Index}
	
	First, we set the initial $K_{\textrm{origin}}$ UAVs to cover users arbitrarily distributed on a 2D plane. It is set by selecting the number of clusters based on the hierarchical clustering with the maximal Silhouette index~\cite{ward,ROUSSEEUW198753}, which reflects the similarity among cluster elements. It is also denoted as $Phase=1$ in Algorithm~\ref{euclid}. The available number of clusters ranges from $2$ to the number of users. However, for computational efficiency, we limit this to $K_{\text{max}}+2$, where $K_{\text{max}}$ is a positive integer. The term ``$+2$" acts as a buffer; this means that we give a margin of an additional increase in the number of clusters. 
	

	\subsection{Ellipse Clustering} 	
	\subsubsection{Cluster Generation and Ellipse Fitting} 
	After choosing $K_{\text{origin}}$, we fit the users into the ellipse with the minimum area, for which we solve
	\begin{equation}
	\begin{array}{rrclcl}
	\displaystyle  \underset{A,~b}{\text {min}} & \multicolumn{3}{l}{\log\det A^{-1}} \\
	\textrm{s.t.} & \left\lVert Ax-b\right\rVert_2 \le 1 ~ (\forall x \in  \mathbb{U}_{\text{cond}}),
	\end{array}
	\label{fitting}
	\end{equation}
	where $A\in\mathbb{R}^{2\times2}$ and $b\in\mathbb{R}^2$ determine the ellipse $\{A,b\}=\{x\in\mathbb{R}^2:\left\Vert Ax-b \right\Vert_2 \le 1$\}. Without loss of generality, we can assume that $A$ is positive definite. $\mathbb{U}_{\text{cond}}$ are the coordinates of the users to be clustered. Initially, it is given by the coordinates of all the users, which is $\mathbb{U}$. Problem ($\ref{fitting}$) implies that we need to find the minimum-area ellipse $\{A,b\}$. It has the volume of $\pi \det A^{-1}$ and covers $\mathbb{U}_{\text{cond}}$.
	
	After dividing $\mathbb{U}_{\text{cond}}$ into $K$ clusters with $\mathbb{U}_1, ..., \mathbb{U}_K$ being the coordinates of users belonging to the respective clusters $1,~...,~K$, we divide cluster $m$ into two sub-clusters, $m_1$ and $m_2$, for all $m\in\{1,~...,~K\}$ through $k$-means clustering. Then, we choose cluster $T$, where its sub-clusters $T_1$ and $T_2$ are ``mostly disjointed." For this, we consider the normalized distance $d_m$ defined as	$d_m =  \frac{\ell_m}{a_m}$, where $\ell_m$ is the centroid distance between sub-clusters $m_1$ and $m_2$, and $a_m$ is the length of the major axis of the ellipse fitted by~(\ref{fitting}) with $\mathbb{U}_{\text{cond}}$ being replaced by $\mathbb{U}_{m}$. From the definition of $d_m$, a larger $d_m$ implies more separation of the corresponding sub-clusters. Hence, we choose $T=\arg\max_{m\in\{1,~...,~K\}} d_m$ and integrate the sub-clusters, except for cluster $T$. This procedure increases the number of clusters from $K$ to $K+1$, where cluster $T$ is split into sub-clusters $T_1$ and $T_2$. Next, we fit their minimum-area ellipses by (\ref{fitting}) with $\mathbb{U}_{\text{cond}}$ being replaced by $\mathbb{U}_{T_i}$: the coordinates of users belonging to sub-cluster $T_i~(i=1,~2)$, and reassign the cluster index $T_1$ as $T$ and $T_2$ as $K+1$. We repeat this process until the number of clusters becomes $K_{\text{origin}}$.
	
	\subsubsection{Ellipse Intersection Removal} 
	
	After assigning the users in $\mathbb{U}_{\text{cond}}$ to $K_{\text{origin}}$ ellipses, we check the intersection between pairs of ellipses by performing a brute-force search that retrieves the indices of intersecting ellipses, i.e., those sharing one or more users. We denote the set of intersection existence indices as $\mathcal{I}$, and update $K_{\text{max}}$ as $K_{\text{max}}=|\mathcal{I}| $. Then, we consider the coordinates $\mathbb{U}_{m}$ with $m\in \mathcal{I}$ and update their unions to $\mathbb{U}_{\text{cond}}=\cup_{m \in \mathcal{I}} \mathbb{U}_{m}$. We repeat the procedure from Section~\RomanNumeralCaps{3}-A and repeat the algorithm on $\mathbb{U}_{\text{cond}}$ until no ellipse intersection remains. By setting $K_{\text{max}}$, which is related to an upper bound for $K_{\text{origin}}$ in Section~\RomanNumeralCaps{3}-A, to $|\mathcal{I}|$, we prevent a considerable increase in the number of clusters.
	
	We should consider the possibility that each cluster is overlapped by another one, i.e., $|\mathcal{I}|=K_{\text{max}}=K_{\text{origin}}$. In this case, we automatically set $K_{\text{origin}}$ as one more than the number of overlapping clusters ($=K_{\text{max}}+1$) to avoid an infinite loop caused by selecting $K_{\text{origin}}$ by the hierarchical clustering and finding that all groups overlap once again. To perform this, we set $Phase$ as ``$Phase=2$”; ``2” has no meaning and is just set to prevent the variable from being ``1”.
	
			\begin{algorithm} [t]
		\caption{Ellipse Clustering}\label{euclid}
		\begin{algorithmic}[1]
			\Procedure{Ellipse Clustering}{}\newline
			\textbf{Input:} $K_{\text{max}}$, Coordinates of users $\mathbb{U}$  \newline
			\textbf{Output:} Number of UAVs $M$, User--UAV association $\{C_m\}_{m=1}^M$, Fitted ellipse $\{A_m,b_m\}_{m=1}^M$ \newline
			\textbf{Initialization} $Phase \leftarrow 1$ , $\mathbb{U}_{\text{cond}}$ $\leftarrow$ $\mathbb{U}$, $M\leftarrow0$
			\While{$\mathbb{U}_{\text{cond}}\neq \emptyset$}
			\If {$Phase=1$}
			\State$K_{\text{origin}}$$\leftarrow$Hierar.\_Sil.($\mathbb{U}_{\text{cond}}$,~$K_{\text{max}}+2$)
			\ElsIf {$Phase=2$}
			\State $K_{\text{origin}}\leftarrow K_{\text{max}}+1$
			\EndIf	      
			\State  $K$ $\leftarrow$ 1, \{$A_1$,$b_1$\}$\leftarrow $Elps.\_Fit.($\mathbb{U}_{\text{cond}}$) 
			\While{$K$ $\neq$ $K_{\text{origin}}$}
			\State Divide every cluster into two sub-clusters
			\State (by $k$-means clustering)
			\State $T\leftarrow\text{argmax}_{m\in\{1,~...,~K\}} d_m$
			\State Integrate sub-clusters except cluster $T$
			\State \{$A_{T_i}$,$b_{T_i}$\}$\leftarrow$Elps.\_Fit.($\mathbb{U}_{T_i}$)~($i=1,~2$)
			\State clear $\mathbb{U}_T$, $T \leftarrow T_1, K+1 \leftarrow T_2, K \leftarrow K+1$
			\EndWhile
			\State $M \leftarrow M+K_{\textrm{origin}}$, [$K_{\text{max}}$, $\mathbb{U}_{\text{cond}}$, $Phase$]
			\State$\leftarrow$Inters.\_Remov.$(\{\mathbb{U}_m\}_{m\in\mathcal{I}}$, $\{A_m,b_m\}_{m\in\mathcal{I}}$)
			\EndWhile
			\EndProcedure
		\end{algorithmic}
	\end{algorithm}
	
	\subsection{Complexity of the Ellipse Clustering Algorithm} 
				The complexity of the initial iteration of Algorithm~\ref{euclid} consists of three parts. First, the complexity of the hierarchical clustering based on the Silhouette index is given by $O(N^2 \log N+N^2)=O(N^2 \log N)$~\cite{ward, ROUSSEEUW198753}. The complexity of solving the equation~(\ref{fitting}) with respect to $\mathbb{U}$ is given by $O(\sqrt {N} \log N)$ by using the interior-point method~\cite{boyd}. In cluster generation step, the complexity is determined by dividing the sub-clusters by the $k$-means algorithm and selecting $T$, such that $\sum_{k=1}^{K} O(|\mathbb{U}_k|)=O(N)$, and fitting the selected sub-clusters with a complexity of $O(\sqrt{|\mathbb{U}_{T_1}|} \log |\mathbb{U}_{T_1}| + \sqrt{|\mathbb{U}_{T_2}|}\log |\mathbb{U}_{T_2}| )\le O(\sqrt{N} \log N)\le O(N)$. Therefore, the complexity of cluster generation is bounded by $O(K_{\textrm{origin}}N)\le O(MN)$. Finally, the complexity of the intersection removal step is performed in a brute-force manner, which is therefore given by $O(K_{\textrm{origin}}^2)\le O(M^2)$. Hence, by considering $I$ iterations until $\mathbb{U}_{\textrm{cond}}\neq\emptyset$, the total complexity is upper-bounded by $O(I(N^2 \log N+MN+M^2))$, which requires far less time and effort than a brute-force search of the optimal association without ICI.
		\begin{figure}[t]
		\begin{center}
			\includegraphics[width=0.632\columnwidth,keepaspectratio]%
			{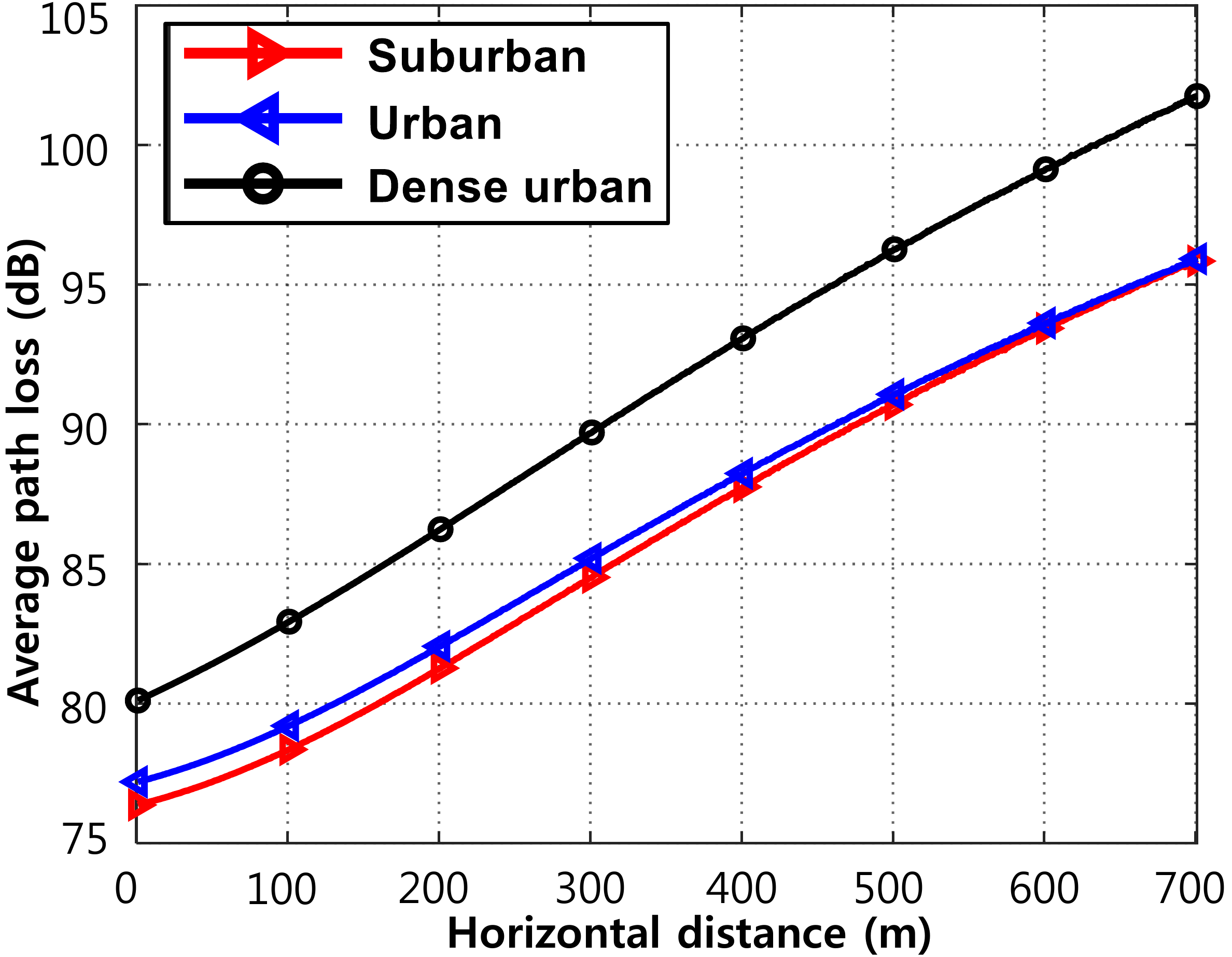}
			\caption{Average path loss according to the horizontal distance between the UAV and the user.}
			\label{Horizion_PL}
		\end{center}
	\end{figure}

	\section{Determining UAV 3D Location} 
	Once the elliptic clusters are formed and $M$ is determined, each UAV $m$ is placed at its cluster center given by $A_m^{-1}b_m$, and its orientation is set according to the major and minor axes of the corresponding ellipse. During the 3D positioning of UAVs, we determine each UAV's altitude and antenna HPBWs. This is done to minimize the transmit power and adjust the orientation of each UAV while avoiding inter-cell interference. Fig.~\ref{Horizion_PL} shows the average path loss according to the horizontal distance user to the corresponding UAV at an altitude of $h=300~\textrm{m}$ under different urban environments. The average path loss of the user increases with an increase in the horizontal distance. Hence, the user with the maximum horizontal distance to its associated UAV experiences the highest average path loss. Consequently, to guarantee the QoS, we only have to guarantee the QoS for the farthest cell-edge user to its associated UAV and relate the decision of the UAV transmit power for it.
	
	Let $P^{(m)}_{r,e}$ and $\overline{{\it PL}}^{(m)}_{e}$ be the received power and average path loss of the farthest cell-edge user from the UAV $m$, respectively. Thus, the minimum required transmit power $P_{\text{req},m}$ for UAV $m$ can be expressed as
	\begin{align}
	P_{\text{req},m} = \mathrm{min} (P_{t,m})= P^{(m)}_{\text{th},e}\cdot \mathrm{min}(\overline{{ PL}}^{(m)}_{e}),
	\label{preq}
	\end{align}
	where $P^{(m)}_{\text{th},e}$ is the received power of the farthest cell-edge user that satisfies the QoS threshold. We can observe that the minimum transmit power is related to the minimization of the average path loss. Therefore, we have to find $h^{\star}$ that minimizes $\overline{{ PL}}^{(m)}_{e}$ under $h_{\textrm{min}}\le h_m \le h_{\textrm{max}}$.
	
\begin{figure}[t]
			\begin{center}
				\includegraphics[width=0.62\columnwidth,keepaspectratio]%
				{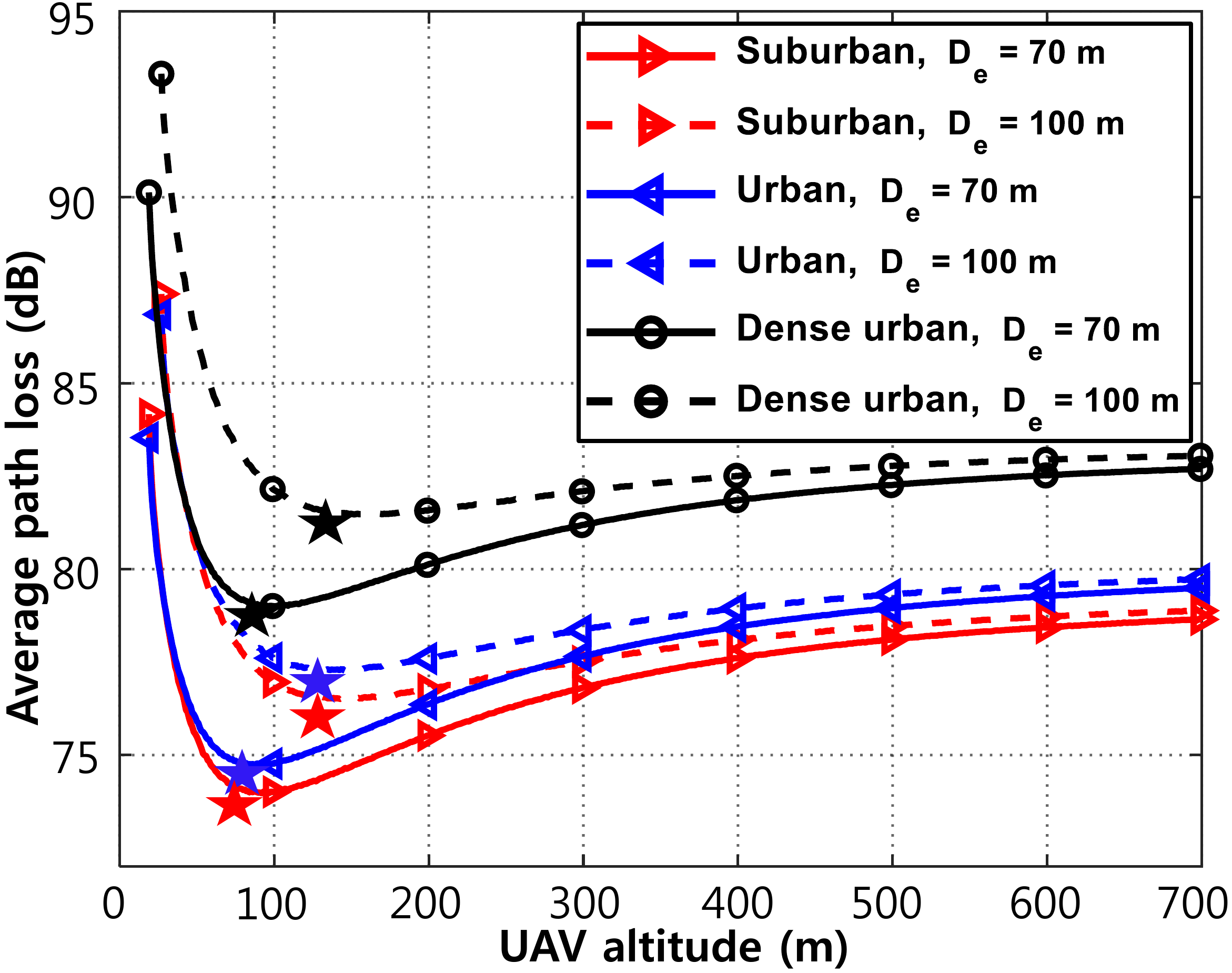}
				\caption{Average path loss according to UAV altitude.}
				\label{Altitude_PL}
			\end{center}
		\end{figure}

	Let $D_{e}$ be the horizontal distance between the UAV and the farthest cell-edge user from it. Fig.~\ref{Altitude_PL} shows the average path loss according to UAV altitude for different environments and horizontal distances. For given $D_e$, we can find the altitude $h^{\star}$ that globally minimizes the average path loss, which implies the quasiconvexity of $\overline{{ PL}}^{(m)}_{e}$~\cite{boyd}. This is because when UAV altitude is too low, the effect of NLoS link dominates the path loss, which results in a sharp increase of average path loss. Moreover, regardless of the environment, increasing $D_{e}$ leads to an increase in the minimum average path loss. 
	Therefore, considering the quasiconvexity of $\overline{{ PL}}^{(m)}_{e}$ in Fig.~\ref{Altitude_PL}, we can find the altitude $H$ of each UAV that minimizes the average path loss as follows~\cite{boyd}:
	\begin{align}
	\label{Altitude}
		H = \left\{ \begin{array}{cl}
		h^{\star} & (h^{\star} \in [h_{\textrm{min}}, h_{\textrm{max}}]) \\
		\textrm{argmin}_{\{h_{\textrm{min}},~h_{\textrm{max}}\}} \overline{{ PL}}_e^{(m)} & (\textrm{otherwise}). \end{array} \right. 
		\end{align} 
		Using this result, finally, we can also finally determine the corresponding HPBWs with $2\Theta_i=2\tan^{-1} (\frac{H}{r_i})~(i=1, 2)$ for each UAV.
				\begin{center}
			\begin{table}[t]
				\caption{Simulation parameters.}
				\begin{tabular}{|>{\centering} p{1.4cm} |>{\centering} p{4.8cm} |>{\centering} p{1.4cm} | }
					\hline
					\textbf{Paramet{\tiny }er} & \textbf{Description} & \textbf{Value}
					\tabularnewline
					\hline
					\centering		$K_{\text{max}}$ & Hierarchical clustering parameter  & 8 \tabularnewline \hline
					\centering			$\zeta_{\text{th}}$  & SNR threshold & 0~dB \tabularnewline \hline
					\centering			\scriptsize	$N_0$  & Noise power spectral density & \text{-}170~dBm/Hz \tabularnewline \hline
					\centering	$\mathcal{E}_{\text{LoS}},~\mathcal{E}_{\text{NLoS}}$ & Additional path loss for LoS, NLoS  & 3,~34~dB \tabularnewline \hline
					\centering		$f_c$ & Carrier frequency & 2 GHz \tabularnewline \hline			
				\end{tabular}
				\label{SimulationParameter}
			\end{table}
		\end{center}
	
	\section{Simulation Results}
	For simulations, we considered 100 independent user distributions following a Poisson Cluster Process (PCP), within a geographical area of 1~km$\times$1~km. In addition, we considered an urban environment in~\cite{a2gglobecom} to model the LoS and NLoS probabilities. Detailed parameters are listed in Table~\ref{SimulationParameter}.
	
\begin{figure}[t]
	\begin{center}
		\includegraphics[width=0.9\columnwidth,keepaspectratio]%
		{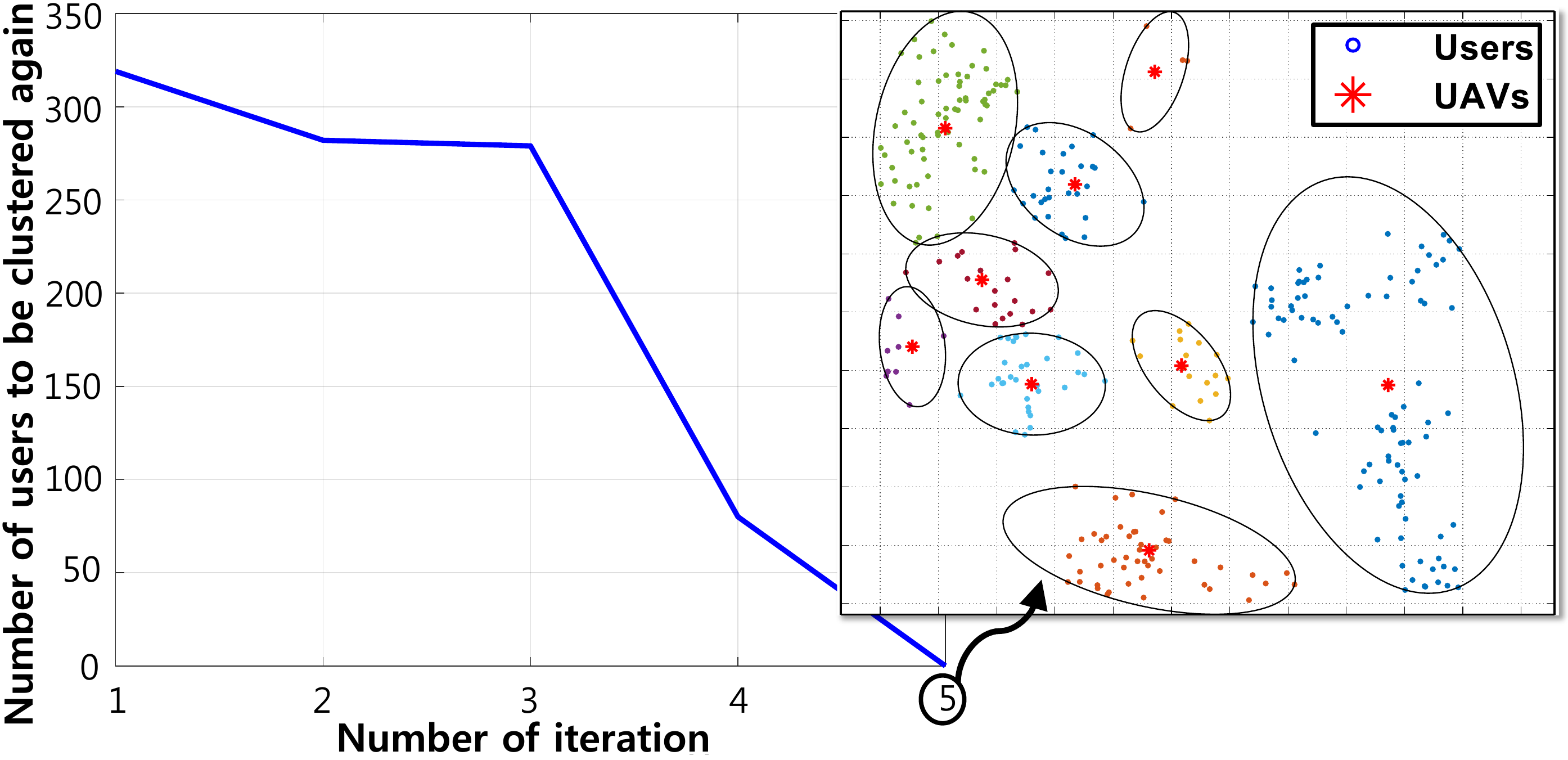}
		\caption{Overall convergence of the algorithm and user--UAV associations based on ellipse clustering.}
		\label{ell}
	\end{center}
\end{figure}

	Fig.~\ref{ell} shows the convergence of the algorithm and the locations of nine UAVs and their associated users by applying the proposed ellipse clustering in one of the user distributions with 319 users. As we can see from the figure, $|\mathbb{U}_{\textrm{cond}}|$ converges after 5 iterations. At the end of the algorithm, the UAVs are located at the center of the corresponding ellipses, and the users covered by each UAV do not experience inter-cell interference with QoS guaranteed. 
	
	For comparison, we considered UAV positioning based on the circle-packing approach proposed in~\cite{mozaeff}. It considers fixed UAV altitude and transmit power to maximize the coverage lifetime while guaranteeing QoS to every user. The coverage is circular, given the fixed azimuth and elevation HPBWs. Table~\ref{bar} lists the coverage probability and the total transmit power of the UAVs using the circle-packing approach and our ellipse clustering. We computed the average total transmit power of more than 100 user distributions by applying the ellipse clustering. As the coverage probability of the circle-packing approach increases, the UAVs require more power. In addition, increasing UAVs does not proportionally increase the coverage probability because of the nonoverlapping condition of circles. In contrast, our ellipse clustering considerably improves the coverage probability compared with the case of fixed altitude and transmit power. Moreover, when using the same number of UAVs with the circle-packing approach, the total transmit power decreases from 44 mW to 22 mW. Note that the subspace brute-force shown in Table~\ref{bar} is based on brute-force search for reasonable search spaces since full brute-force is infeasible due to its computational complexity.

\begin{figure}[t]
		\begin{center}
			\includegraphics[width=0.62\columnwidth,keepaspectratio]%
			{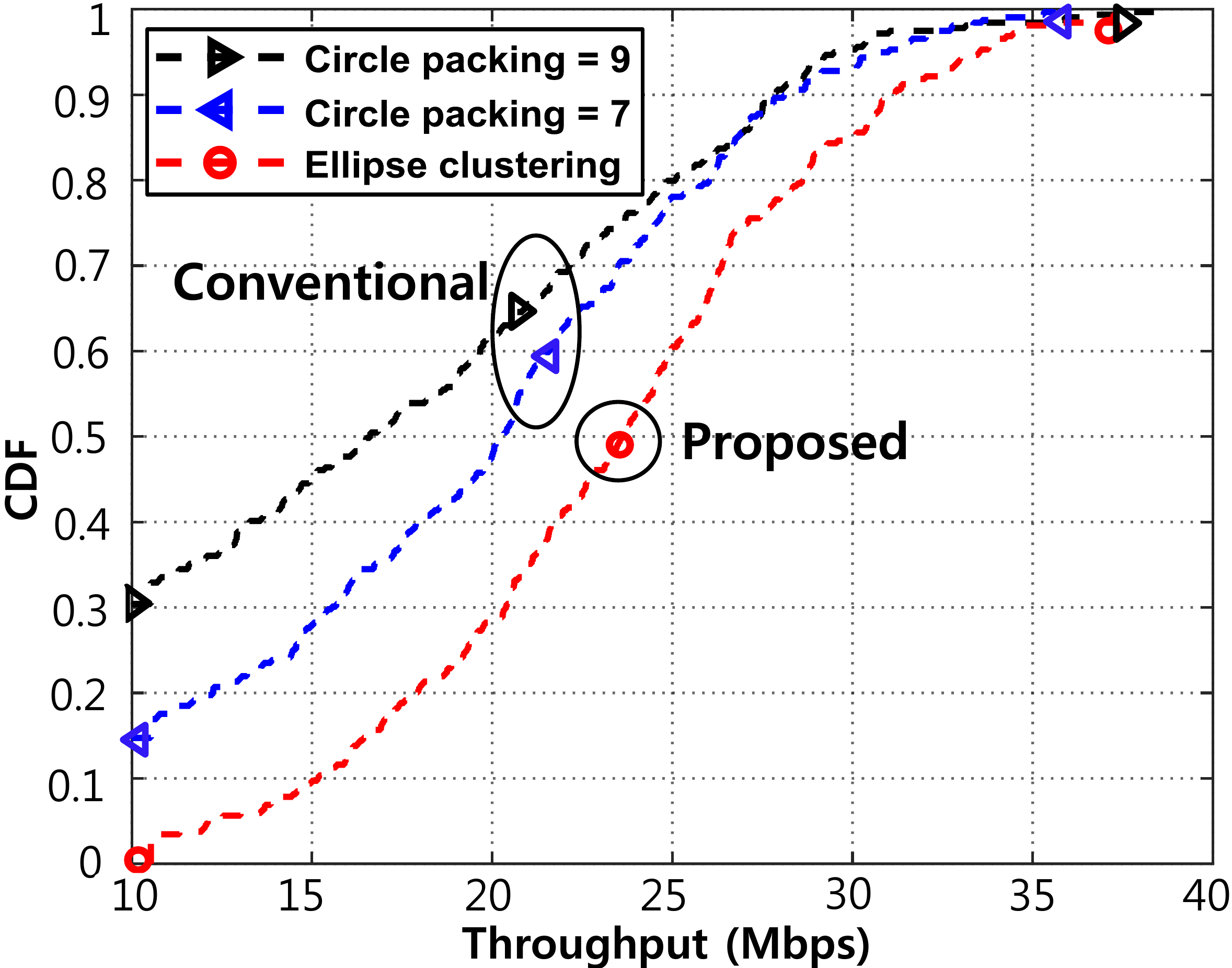}
			\caption{Average throughput cumulative distribution function of users in an urban environment.}
			\label{sls}
		\end{center}
	\end{figure}

	Fig.~\ref{sls} compares the system-level performance of the evaluated coverage methods. The ellipse clustering outperforms the circle-packing approach. This is because the ellipse clustering associates the 3D location of UAV concerning the users' positions to achieve high packing density, while the circle-packing approach is based on fixed altitude and transmit power that neglects the users' positions.

	\section{Conclusion}
    We proposed a framework for efficiently deploying multiple UAVs using ellipse clustering. To provide communication coverage to a target area, operators determine the UAV's azimuth, elevation HPBWs, and orientation. Furthermore, the optimal altitude of each UAV minimizes the total transmit power while guaranteeing the QoS to every user. The simulation results of the proposed method confirm that the total transmit power of UAVs significantly decreases, compared with the circle-packing approach, further guaranteeing the coverage of every user by a UAV and providing higher throughput. 
	
	\begin{center}
		\begin{table}[!t]
			\caption{Total transmit power and coverage probability.}
			\begin{tabular}{|>{\centering} p{4cm} |>{\centering} p{2.1cm} |>{\centering} p{1.4cm} | }
				\hline
				\textbf{Clustering method \\ / the number of UAVs} & \textbf{Total transmit power of UAVs} & \textbf{Coverage probability}
				\tabularnewline
				\hline
				\centering			Circle packing~\cite{mozaeff} / 7 & 52~mW & 0.85 \tabularnewline \hline
				\centering			Circle packing~\cite{mozaeff} / 9  & 44~mW & 0.69  \tabularnewline \hline
				\centering		 \textbf{(Proposed) Ellipse clustering / 9} & \textbf{22~mW} & \textbf{1} \tabularnewline \hline
				\centering			Subspace brute-force / 9 & 18~mW & 1 \tabularnewline \hline
			\end{tabular}
			\label{bar}
		\end{table}
	\end{center}

	
	
	\bibliographystyle{IEEEtran}
\bibliography{UAV_ref}	
	

\end{document}